\documentclass[prl,twocolumn,showpacs,superscriptaddress]{revtex4}

\usepackage{graphicx,bm,amssymb,amsmath}
\begin{document}

\title{Local contact numbers in two dimensional packings of frictional disks}

\author{Silke Henkes}
\affiliation{Instituut-Lorentz, Leiden University, P. O. Box 9506,
2300 RA Leiden}
\author{Kostya Shundyak}
\affiliation{Instituut-Lorentz, Leiden University, P. O. Box 9506,
2300 RA Leiden}
\author{Wim van Saarloos}
\affiliation{Instituut-Lorentz, Leiden University, P. O. Box 9506,
2300 RA Leiden}
\author{Martin van Hecke}
\affiliation{Kamerlingh Onnes Lab, Leiden University, P. O. Box
9504, 2300 RA Leiden}

\date{\today}

\begin{abstract}
We analyze the local structure of  two dimensional packings of
frictional disks numerically. We focus on the fractions $x_i$ of
particles that are in contact with $i$ neighbors, and
systematically vary the confining pressure $p$ and friction
coefficient $\mu$. We find that for all $\mu$, the fractions $x_i$
exhibit powerlaw scaling with $p$, which allows us to obtain an
accurate estimate for $x_i$ at zero pressure. We uncover how these
zero pressure fractions $x_i$ vary with $\mu$, and introduce a
simple model that captures most of this variation. We also probe the 
correlations between the contact numbers of neighboring 
particles.
\end{abstract}

 \maketitle

While soft frictionless spheres experience a critical jamming
transition in the limit of zero pressure, where properties such as
elastic moduli, contact number, density, characteristic
frequencies and lengthscales exhibit powerlaw scaling
\cite{O'Hern2003,mu0dos,ellenbroek,martinsjammingreview}, the
situation is more delicate for frictional systems. The approach to
the jamming transition is still governed by the pressure, $p$, but
a range of densities and packing properties can exist depending on
the value of the friction coefficient $\mu$, the mobilization
(ratio of frictional to normal forces) of the frictional contacts
and the packing history \cite{makse,dos,kostya,silkepreprint}. In
particular, in $d$ dimensions, the contact number at jamming,
$z_c$, can take on a range of values between $d+1$ and $2d$, in
contrast to frictionless sphere packings which always reach their
respective isostatic contact number $z_{iso}^0=2d$ at jamming. The
proximity to the isostatic contact number governs the
scaling near jamming --- for frictionless spheres, properties such
as elastic moduli scale with distance to jamming. However, for
frictional packings these properties only scale with distance to
the isostatic limit $z_{iso}^{\mu}=d+1$, and in general {\em not}
with distance to jamming \cite{martinsjammingreview,dos,kostya},
although this depends on whether fully mobilized contacts are treated
as frictional or slipping \cite{silkepreprint}.

We recently studied the case of frictional spherical disks in two
dimensions, and focussed on packings that were equilibrated very
gently \cite{dos,kostya,silkepreprint}. This eliminates
preparation history and mobilization as unknowns: for given
pressure $p$ and friction coefficient $\mu$, packings with well
defined statistics are obtained. The gentle equilibration
procedure also allows to approach the isostatic limit for
frictional systems, $z_c=z_{iso}^{\mu}=d+1$ when $\mu \rightarrow
\infty$ and $p\rightarrow  0$ --- here jamming has many of the
critical features observed for frictionless systems
\cite{dos,kostya}.


One additional surprise is that for finite values of $\mu$, such
gently equilibrated packings still reach a {\em generalized}
isostatic limit \cite{kostya,silkepreprint}. This means, in short,
that a substantial number of contacts get fully mobilized, i.e.,
their frictional forces $f_t$ satisfy the bound $|f_t|\le f_n$,
where $f_n$ denotes the normal force. If these fully mobilized
contacts are seen as slipping, the critical nature of the
vibrational density of states at jamming is restored for all
values of $\mu$ \cite{silkepreprint}.

\begin{figure}[tb]
\includegraphics*[width=8.6cm]{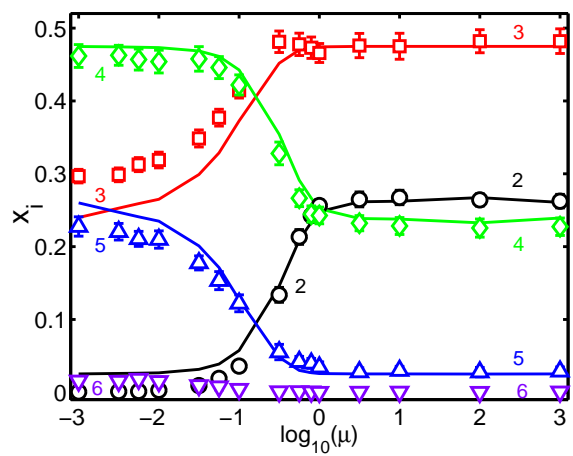}
\caption{\label{xmup0}{Variation of the fractions $x_i(p=0,\mu)$
of particles with $i=2,3,\ldots,6$ contact neighbors as function
of the friction coefficient $\mu$. The full curves are predictions
from a simple model (Eqs.~\ref{eq:xsolution}) with fixed variance
$\sigma^2=0.6$. }}
\end{figure}

Here we probe the fractions $x_i(p)$ of particles that have $i$
contacts for these frictional packings. These fractions are the
simplest characteristics of the contact network beyond the average
contact number $z$. It is thus natural to ask how the fractions
$x_i$ depend on $p$ and $\mu$. We find that, for given $\mu$, the
fractions $x_i(p)$ exhibit scaling with $p$ similar to the scaling
of the total contact number $z$. This allows us to extrapolate
these fractions to $p\rightarrow0$, and this is the case on which
we focus our attention. As is shown in Fig.~\ref{xmup0}, the
fractions $x_i$ vary substantially with $\mu$, and reach
well-defined values in the limits where $\mu\rightarrow 0$ or
$\mu\rightarrow \infty$. We find a number of simple but unexpected
relations between the various $x_i$, and introduce a simple model
that, given $z(\mu)$, gives a good prediction for $x_i(\mu)$.

{\em Packings} --- Following \cite{Somfai2005}, the numerical
systems under consideration are two dimensional packings of $1000$
spheres with $20\%$ polydispersity in the diameter of the
particles in a square box with periodic boundary conditions.
The grains interact through 3$d$ Hertz-Mindlin forces, i.e with the normal
force $f_{ij}$ between particles $i$ and $j$ proportional to
$\delta^{3/2}_{ij}$, with $\delta_{ij}$ the overlap of the
two particles.
The Young modulus of the grains is set to 1, which determines the
pressure unit, and the Poisson ratio is set to zero, while the
unit of length is the average grain diameter. The construction and
equilibration of the packings has been described in detail
elsewhere~\cite{Somfai2005,kostya}. Rattlers, particles which have
no appreciable interactions with any of the other particles, are
always left out of the analysis of the packings and contact
statistics. For each value of $\mu \in[10^{-3},10^3]$ and
$p\in(10^{-6},10^{-3})$, 30 configurations were generated
independently.

\begin{figure}[tb]
\centering
\includegraphics[width = 0.99\columnwidth, trim = 0mm 0mm 0mm 0mm, clip]{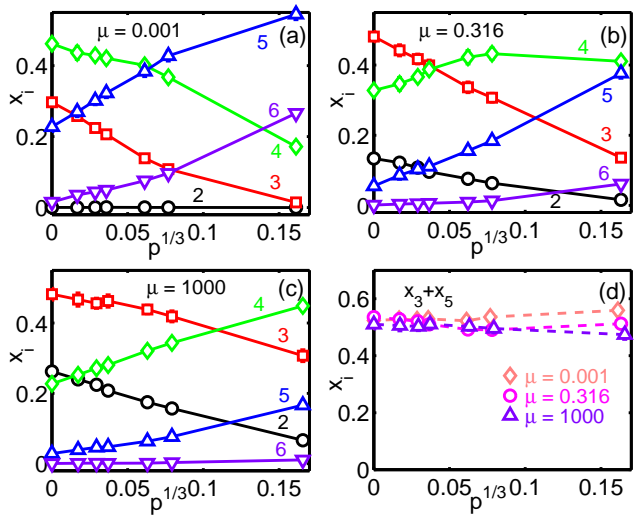}
\caption{Contact fractions $x_{i}$ as a function of pressure.
(a-c) For three representative values of $\mu$, the $x_{i}$ scale
linearly with $p^{1/3}$ (equivalent to $\phi^{1/2}$ for the
hertzian interaction), and we are able to extrapolate to $p=0$.
(d) The sum $x_{3}+x_{5}\approx 0.5$, for all values of $\mu$ and
$p$ studied.} \label{fig:ppanel}
\end{figure}

{\em Scaling of fractions $x_i$ with pressure ---} As is shown in
Fig.~\ref{fig:ppanel}a-c,  $x_i(p,\mu)$ scales linearly with
$p^{1/3}$, which allows us to extrapolate their values for finite
$p$ to the (un)jamming limit at $p=0$. This scaling is the same as
the scaling of the total contact number $z$ with $p$, which for
the Hertzian interactions employed here is consistent with the
scaling that the excess contact number $\Delta z := z-z_c$ scales
with the square-root of the excess packing fraction. This relation
is well known for frictionless systems \cite{O'Hern2003,durian},
but also appears to hold for frictional systems
\cite{dos,leopreprint}
--- our data here suggests that it also holds for the individual
contact fractions, irrespective of the value of $\mu$.

A second robust finding is illustrated in Fig.~\ref{fig:ppanel}d:
the number of particles that have an odd number of contacts
\cite{footnote35}, is close to 1/2 --- the number of particles
with an even or odd number of contacts is therefore approximately
equal, irrespective of pressure or value of $\mu$. We do not have
a satisfactory explanation for this.

{\em The extrapolated fractions $x_i$ at jamming ---} In the
remainder of this paper we focus on $x_i(\mu)$ at zero pressure.
Since $x_i$ has to be zero for $i=1$, and the fraction of
particles with $7$ contact are negligible for the polydispersities
employed here we focus on $i$ ranging from 2 to 6. As shown in
Fig.~\ref{xmup0}, the variation of $x_i$ with $\mu$ is greatest
for $\mu$ between 0.1 and 1, with the small and large $\mu$ limits
apparently well behaved.

The functional forms of $x_i(\mu)$ for $i$=3 and 5 are similar, as
are the functional forms of $x_i(\mu)$ for $i$=2 and 4. This is
related to the observations that $x_3+x_5\approx1/2$. One also
notices that,  approximately, $x_n(\mu\rightarrow 0) \approx
x_{n+1} (\mu \rightarrow \infty)$. In fact, for small $\mu$, the
fractions $x_3$ and $x_5$ tend to 1/4, while $x_4$ approaches 1/2
--- for large $\mu$, $x_2$ and $x_4$ tend to 1/4, while $x_3$ approaches
1/2.

In the limits $\mu=0$ or $\mu = \infty$, we can estimate
these fractions by a very simple argument. Let us first focus on
the zero friction case. Assuming that there are only particles
with three, four or five contacts, the fractions $x_3,x_4$ and $
x_5$ can immediately be calculated, since combining the condition
that $x_3+x_4+x_5=1$ with the isostaticity condition
$3x_3+4x_4+5x_5=4$ implies $x_3=x_5$, and hence $x_3=1/4, x_4=1/2$
and $x_5=1/4$ --- a similar argument holds for $x_2,x_3$ and $
x_4$ in the limit of infinite friction. Deviations from this result
 arise since a small fraction of particles with respectively
six and five contacts arise, weakly breaking the ``three particle
species '' condition underlying this argument (see Fig.~1).

\begin{figure}[t]
\centering
\includegraphics[width = 0.9\columnwidth, trim = 0mm 10mm 10mm 0mm, clip]{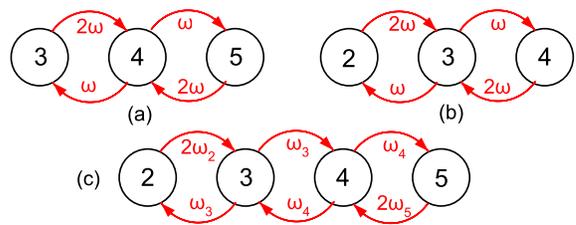}
\caption{Rate equation models for the equilibrium contact
fractions. (a,b) A model with a single rate $\omega$ is sufficient
for $\mu\rightarrow 0$ and $\mu\rightarrow 0$. (c) For finite
$\mu$, we introduce individual rates $\omega_{i}$ which correspond
to the total decay rate for contact number $i$.}
\label{fig:ratemodel}
\end{figure}

{\em Simple rate equation model ---}
The ratios $x_3/x_4=x_5/x_4=1/2$ can also be understood in terms
of a simple stochastic model where we imagine distorting a certain
packing, creating and breaking contacts but keeping the overall
contact number and the ratios $x_i$ constant. In the case of three
species only, particles with 4 contacts can become 3's and 5's,
while 3's and 5's can only become 4's (See Fig.~3). Since the
transition probabilities must all be equal (since always two
particles take place in such an event), and, on average, we
require the fractions $x_i$ to be constant, we get, in this simple
approximation, $x_4=2x_3=2x_5$. This heuristic argument can be
written as a rate equation model, as shown in
Figure~\ref{fig:ratemodel}a. Once we normalize the rates such that
the \emph{total} decay rate of each species is $2\omega$, we
obtain as steady state $x_4=2x_3=2x_5$.

For intermediate values of $\mu$, the number of species is four
(if we neglect a small number of $z=6$-contacts). A single decay
rate would than imply that $\lbrace x_{2}\approx 1/6, x_{3}\approx
1/3, x_{4}\approx 1/3,  x_{5}\approx 1/6 \rbrace$ and $z=3.5$ ---
clearly a single rate does not capture the data.
Figure~\ref{fig:ratemodel}c shows an extended model where we now
associate an individual rate $\omega_{i}$ to each species $i$, so
that the total decay rate of that species is $2\omega_{i}$. The
solution to this model is $x_{i} \sim 1/\omega_{i}$ for $i=3,4$
and $x_{i}\sim 1/(2\omega_{i})$ for $i=2,5$.

{\em Explicit solutions of rate equation model ---} We now seek an
explicit solution of the four species model for the contact
fractions as a function of the friction coefficient. To achieve
this, we introduce two constraints on the model beyond the trivial
normalization constraints $\sum_{i=2}^{5} x_{i}=1$ and
$\sum_{i=2}^{5} i x_{i} = z(\mu)$. First, we constrain our model
by the empirical observation that the number of particles with odd
and even contacts is equal, i.e., $x_{3}+x_{5}=0.5$. Additionally,
 we impose the variance of the contact fraction distribution, 
$\sum_{i=2}^{5} x_{i} (z-i)^{2} = \sigma^{2}$. The solution
to the resulting set of equations is
\begin{align}
& x_{2}=\left((z-4)^{2}+\sigma^{2}-1/2\right)/4 \nonumber \\
& x_{3}=\left(-(z-3)^{2}-\sigma^{2}+5/2\right)/4  \label{eq:xsolution} \\
& x_{4}=\left(-(z-4)^{2}-\sigma^{2}+5/2\right)/4 \nonumber \\
& x_{5}=\left((z-3)^{2}+\sigma^{2}-1/2\right)/4 \nonumber
\end{align}

\begin{figure}[tb]
\centering
\includegraphics[width = 0.9\columnwidth, trim = 0mm 0mm 10mm 0mm, clip]
{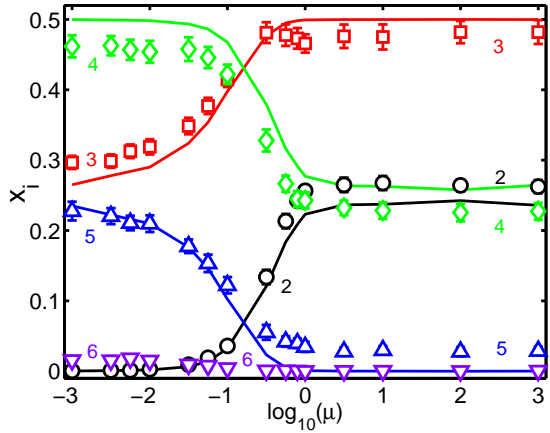} \caption{Contact
fractions as a function of $\mu$ in the extrapolated limit
$p\rightarrow 0$. The curves show the model solution from
equation~\ref{eq:xsolution}, with a variance $\sigma^{2}=0.5$.}
\label{fig:simfractions_sigma0p5}
\end{figure}

To obtain definite predictions from this set of equations, we need
to determine the variance $\sigma^2$. In the extreme limits, and
under the simplifying assumption that only three species with
fractions $1/4,1/2,1/4$ arise, we find $\sigma^2=0.5$ (notice if
more species are present, $\sigma^2$ will be larger). Fixing now
$\sigma^2=0.5$ over the whole range of friction coefficient, we
obtain the prediction shown in
figure~\ref{fig:simfractions_sigma0p5}. There are no additional
fit parameters to this solution, and the agreement is quite good.

We have numerically studied the actual variance of $\sigma^2$ from
the data, and find that for our data it varies between 0.57 and
0.65 --- when we fix $\sigma^2=0.6$, the fit becomes significantly
improved, as shown in Fig.~1.

\begin{figure*}[tb]
\centering
\includegraphics[width = 0.9\columnwidth, trim = 0mm 0mm 10mm 0mm, clip]{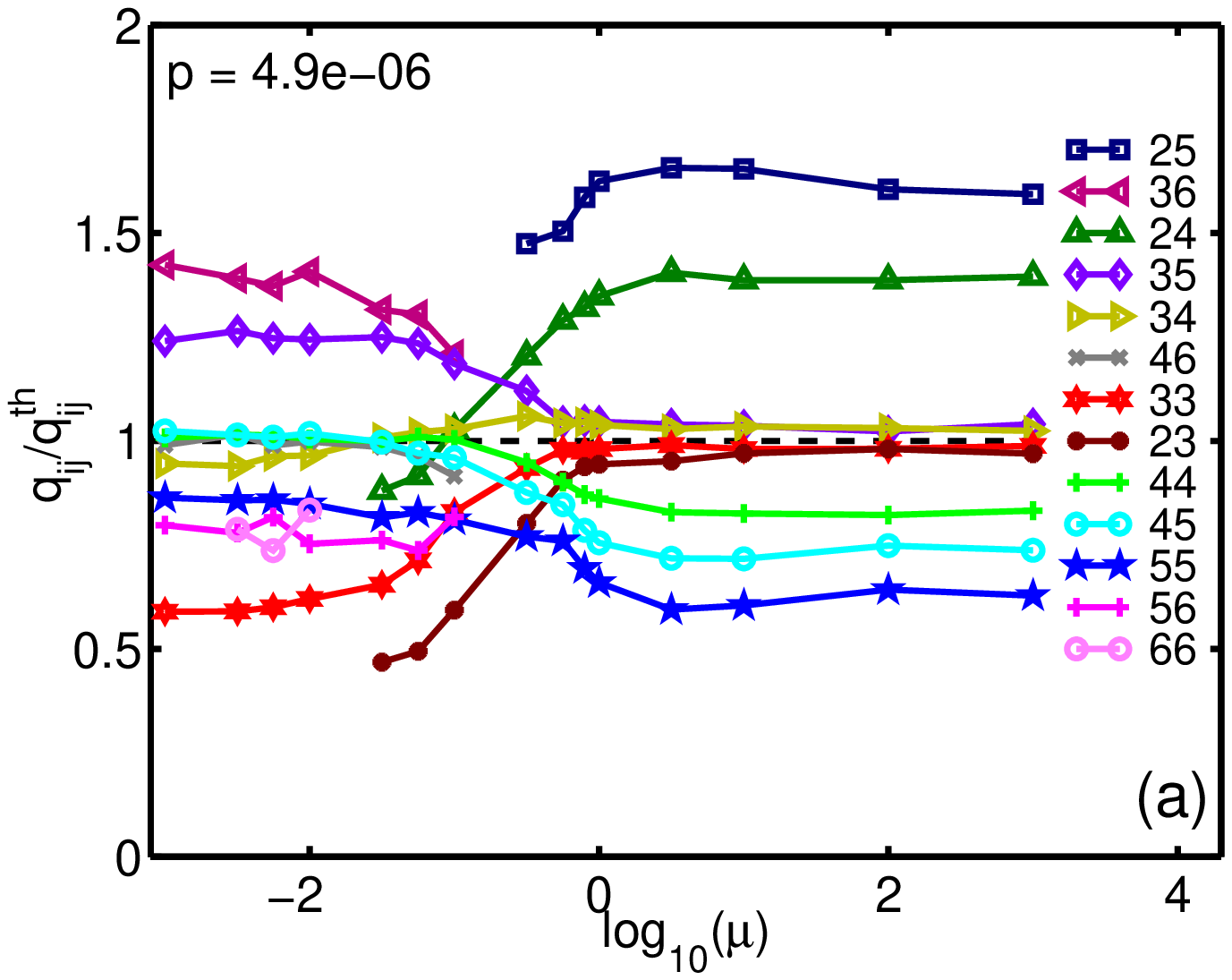}
\includegraphics[width = 0.9\columnwidth, trim = 0mm 0mm 10mm 0mm, clip]{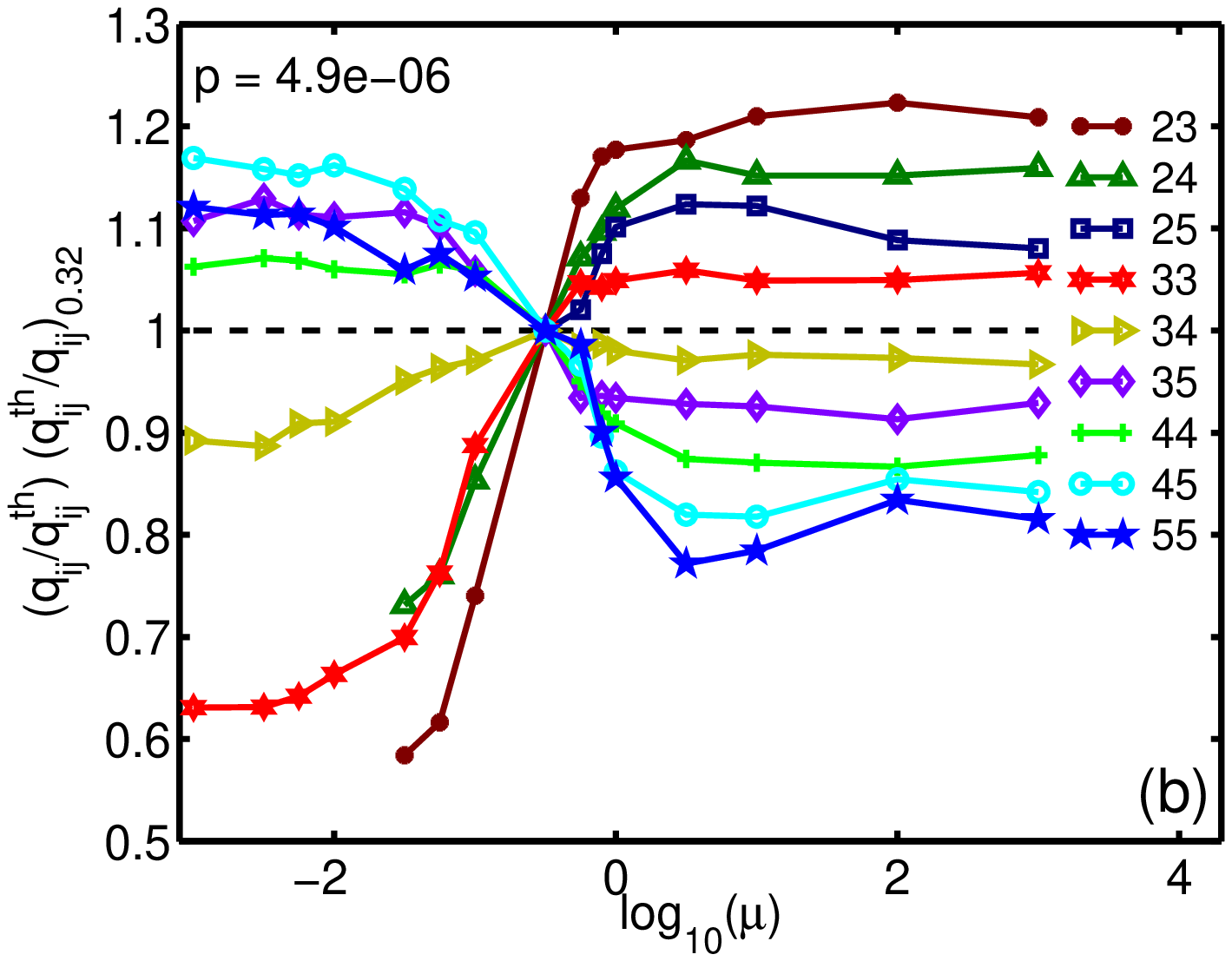}
\caption{(a) Ratio of the observed contact pair fraction $q_{ij}$
to the prediction $q_{ij}^{th}$ from equation \ref{eq:qijtheory}
for all contact pairs with sufficient statistics. Contact pairs
with very dissimilar $i$ and $j$ are favored. (b) Ratio of the
observed contact pair fraction $q_{ij}$ to the prediction
$q_{ij}^{th}$, rescaled by the ratio at $\mu=0.32$. Contact pairs
with large mean contact number reduce in frequency as $z$ drops,
while pairs with small mean contact number show an upward trend.}
\label{fig:qijratios}
\end{figure*}

{\em Correlations ---} The rate equation model derives from its
implicit assumption that the contact numbers of particles and
their neighbors are uncorrelated. Based on this assumption, we can
calculate the theoretical fraction $q_{ij}^{th}$ of contacts
between particles with $i$ and $j$, given $x_i$ and $x_j$. Since
the total fraction of contacts for particles with $i$ contacts is
given by $i x_{i}/z$, the uncorrelated prediction for $q_{ij}$ is
\begin{equation} q_{ij}^{th} = \frac{2 i j x_{i} x_{j}}{z^{2}}
\:\:\text{for}\:\: i\neq j; \quad q_{ij}^{th} = \frac{i j x_{i}
x_{j}}{z^{2}} \:\:\text{for}\:\: i= j \label{eq:qijtheory}
\end{equation}

Figure~\ref{fig:qijratios}a shows the ratio $q_{ij}/q_{ij}^{th}$
of the observed fraction of contacts and the uncorrelated
prediction \cite{footnotep0}. For intermediate values of $i$ and
$j$ the prediction is quite reasonable, as $q_{ij}/q_{ij}^{th}$
remains bounded between 0.5 and 1.6 orso. Contact pairs with very
\emph{dissimilar} $i$ and $j$ are favored - this is likely an
effect of polydispersity, since small particles with few contacts
prefer to sit next to larger particles with more contacts. A
detailed study of this is left for the future.

In Figure~\ref{fig:qijratios}b we have divided out the ratio at an
intermediate $\mu$, to more clearly see the variation of $q_{ij}$
with $\mu$. This shows that the fractions corresponding to
particles with $x_i$ that are abundant (such as $q_{44}$ for small
$\mu$ and $q_{33}$ for large $\mu$) do not vary strongly with
$\mu$. There appears to be a correlation between the relative over
representation of contacts and the overabundance of the species of
particles (i.e., for large $\mu$, there are many particles with 2
or 3 contacts, and $q_{23}$ is over abundant, while there are very
few particles with 4 and 5 contacts, and the ratios $q_{44}$,
$q_{45}$ and $q_{55}$ are even less likely) --- we have no clear
explanation for this.

{\em Outlook ---} Simple arguments allow us to estimate the
contact fractions $x_i$, which can be seen as fingerprints of the
system. Since frictional systems depend on history, we expect the
fractions and their variation to be a useful step in identifying
the effects of preparation history beyond average values such as
overall contact number and density.

We are grateful to W. Ellenbroek and L. Silbert for illuminating
discussions. SH and  KS acknowledge support from the physics
Foundation FOM.


\begin{thebibliography}{10}
\bibitem{footnote35}Note that we approximate the number of particles wiht
an odd number of contacts as $x_3+x_5$. 
This is a good approximation as there are no particles
with one contact, and very few with seven or more
--- for the largest pressures, $x_7=2.5\times 10^{-3}$, and $x_7$
rapidly approaches zero for smaller pressures.
\bibitem{footnotep0} Notice that we use here the values at the
lowest pressure --- extrapolating all these joint fractions for
zero pressure leads to quite large error bars, since some joint
fractions are very small.
\end{thebibliography}
\end{document}